\documentclass[10pt,a4paper]{article}
\usepackage[latin1]{inputenc}
\usepackage[dvips]{epsfig}
\usepackage[francais]{babel}
\usepackage{amssymb,amsmath,amscd,times}
\usepackage{a4wide,graphicx,alltt}
\newcommand{\dummy}{\mbox{}}

\newcommand{\ca}{automate cellulaire }
\newcommand{\cas}{automates cellulaires }
\newcommand{\field}{\mbox{$\mathbb{F}$}}
\newcommand{\ssi}{si et seulement si }
\newtheorem{definition}
        {\bf Définition}

\newtheorem{theorem}
        {\bf Théorème}

\newtheorem{lemma}
        {\bf Lemme}

\begin{document}
\title{Analyse des suites aléatoires engendrées par des automates
  cellulaires et applications à la cryptographie}
\author{Bruno Martin\\
Université de Nice--Sophia Antipolis,\\
Laboratoire I3S, UMR 6070 CNRS,\\
2000 route des lucioles, BP 121,\\
 F-06903 Sophia Antipolis Cedex}
\date{24 mars 2007}
\maketitle
\begin{abstract}
  On s'intéresse aux interactions entre la cryptologie et les
  automates cellulaires. Il a été montré récemment qu'il n'existe pas
  de règle élémentaire d'automate cellulaire non-linéaire robuste à la
  corrélation. Ce résultat limite fortement l'usage d'automates
  cellulaires pour la construction de suites pseudo-aléatoires servant
  de clés utilisables en cryptographie à clé secrète. De plus, pour de
  tels mécanismes de génération de suites pseudo-aléatoires, Meier et
  Staffelbach ont proposé une technique de cryptanalyse
  efficace. Cependant, des pistes subsistent pour construire des
  automates cellulaires susceptibles d'engendrer de bonnes suites
  pseudo-aléatoires, que nous évoquerons à la fin de cet
  article.\\[2ex]
 {\bf Abstract:}
  This paper considers interactions between cellular automata and
  cryptology. It is known that non-linear elementary rule which is
  correlation-immune don't exist. This results limits the use of
  cellular automata as pseudo-random generators suitable for
  cryptographic applications. In addition, for this kind of
  pseudo-random generators, a successful cryptanalysis was proposed by
  Meier and Staffelbach. However, other ways to design cellular
  automata capable to generate good pseudo-random sequences remain and
  will be discussed in the end of this article.
\end{abstract}

\section{Les automates cellulaires}
Les automates cellulaires (ou AC) ont été inventés par Ulam et von
Neumann~\cite{vonne66}. Il s'agit à la fois d'un modèle de système
dynamique discret et d'un modèle de calcul. Un automate cellulaire est
composé d'un ensemble bi-infini de cellules identiques qui peuvent
prendre à un instant donné un état à valeurs dans un ensemble fini. Le
temps est également discret et l'état d'une cellule au temps $t$ est
fonction de l'état au temps $t-1$ d'un nombre fini de cellules appelé
son «voisinage». À chaque nouvelle unité de temps, les mêmes règles
sont appliquées à l'ensemble des cellules, produisant une nouvelle
«configuration» de cellules dépendant entièrement de la configuration
précédente. Nous nous restreindrons ici à des automates cellulaires
sur un anneau de $N$ cellules et dont l'ensemble des états est
binaire.
\begin{definition}
  Un \emph{automate cellulaire} est un ensemble fini de cellules
  identiques indicées par $\mathbb{Z}_N$. Chaque cellule est une
  machine d'états finis $C=(Q,f)$ où $Q=\mathbb{F}_2$ et $f$ une
  application $f:Q\times Q\times Q\rightarrow Q$.
\end{definition}
Le fonctionnement de $f$, la \emph{fonction de transition}, est
le suivant: l'état de la cellule $i$ au temps $t+1$ (noté $x_i^{t+1}$)
dépend des états des cellules $i-1,i$ et $i+1$ au temps $t$ (le
\emph{voisinage} de la cellule $i$ de \emph{rayon} 1):
\[x_i^{t+1}=f(x_{i-1}^t,x_{i}^t,x_{i+1}^t)\]

% \begin{figure}[ht]
%   \centering
%   \includegraphics{UneTrans.pdf}
%   \caption{Transition de la cellule 3 par la règle 30.}
%   \label{fig:UneTrans}
% \end{figure}
\begin{figure}[tbh]
\begin{center}
\begin{minipage}{0.4\textwidth}
  \centerline{\includegraphics[scale=.2]{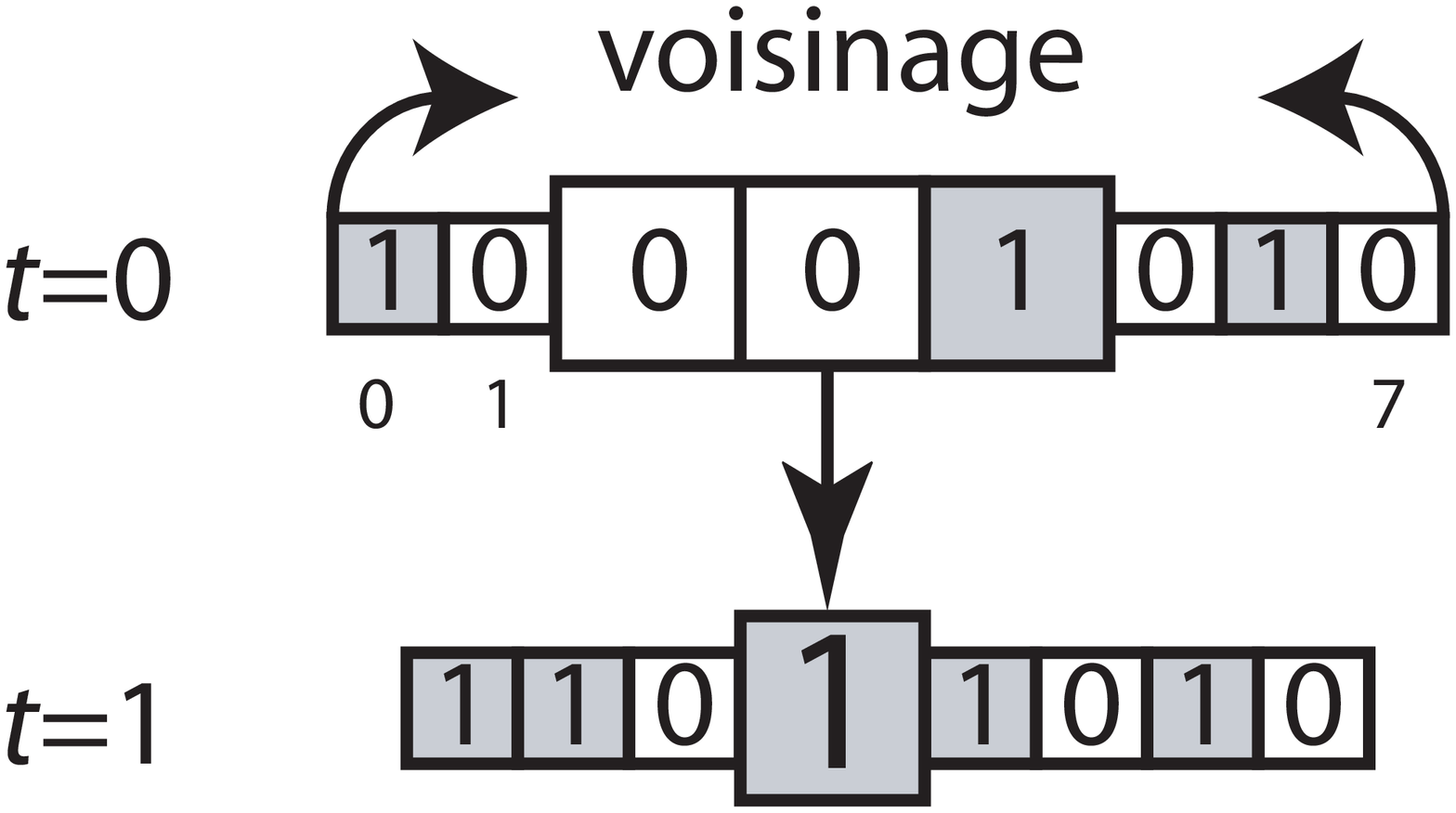}}
  \caption{Transition de la cellule 3.}
  \label{fig:UneTrans}
\end{minipage}
\begin{minipage}{0.4\textwidth}
  \centerline{\includegraphics[scale=.2]{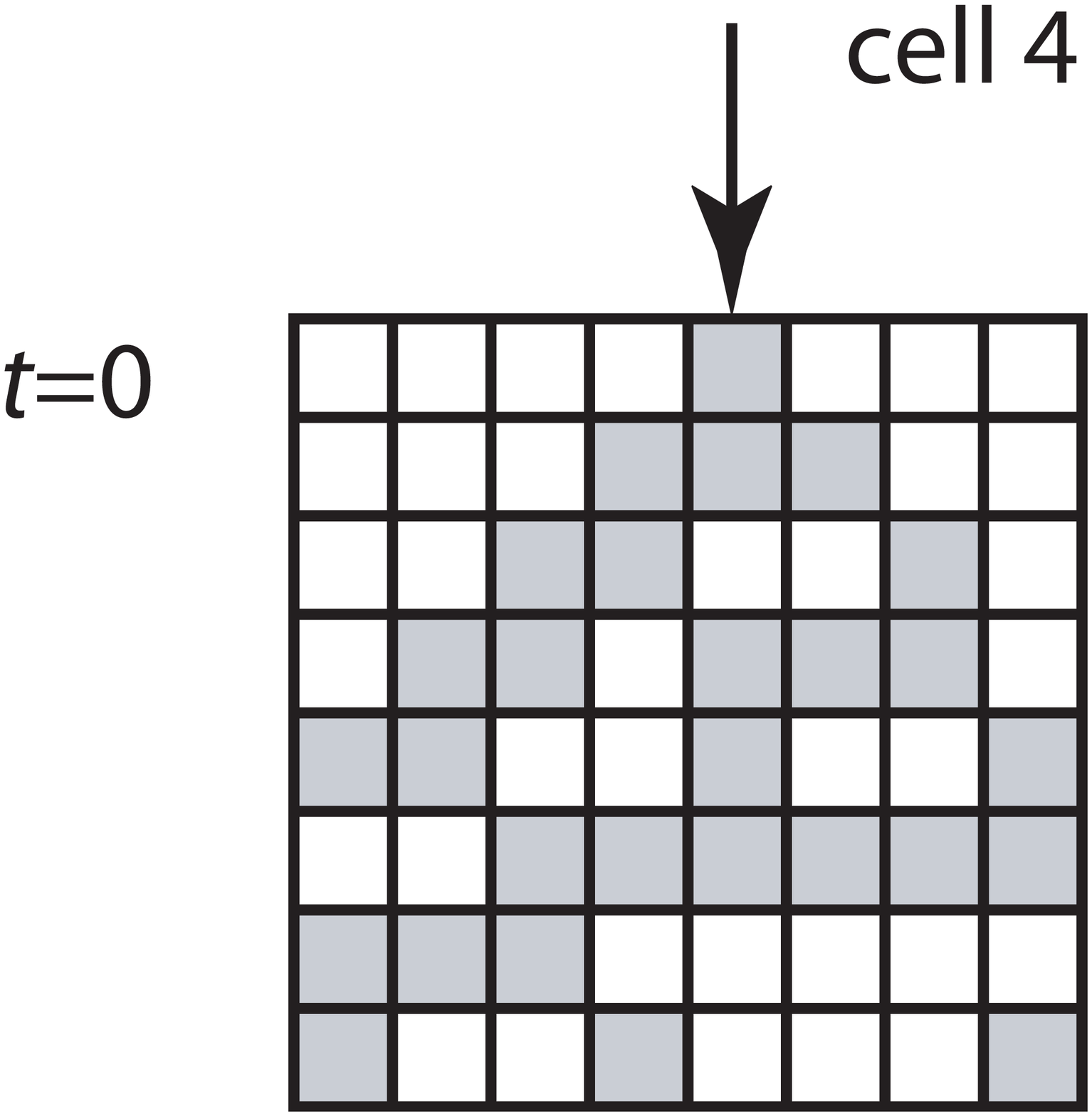} }
   \caption{Evolution d'un AC.}
   \label{fig:CA30PRG}
\end{minipage}
\end{center}
\vspace*{-.8cm}
\end{figure}

La fig.~\ref{fig:UneTrans} illustre une transition d'un automate
cellulaire sur un anneau de 8 cellules. Chacune des cellules pouvant
prendre deux états, il existe $2^3=8$ configurations possibles d'un
tel voisinage. Pour que l'automate cellulaire fonctionne, il faut
définir quel doit être l'état, à l'instant suivant, d'une cellule pour
chacun de ces motifs. Il y a $2^8=256$ façons différentes de s'y
prendre, et donc 256 règles d'automates cellulaires différentes.
Considérons la règle d'automate cellulaire définie par la table suivante:

\[\begin{array}{c|cccccccc}
(x_{i-1}^t x_{i}^t x_{i+1}^t)&111 &110 &101 &100 &011 &010 &001
&000\\\hline
x_i^{t+1}&0 &0 &0 &1 &1 &1 &1 &0 %30
\end{array}\]
(cela signifie que si, à un temps $t$ donné,
une cellule est à l'état «0», sa voisine de gauche à l'état
«0» et sa voisine de droite à l'état «1», au temps $t+1$ elle sera à
l'état «1».)
Si l'on part d'une configuration initiale où toutes les cellules sont
à l'état «0» sauf une, on aboutit à la suite de configurations
(appelée aussi \emph{diagramme espace-temps}) décrit dans la
figure~\ref{fig:CA30PRG}.
% \begin{figure}[h]
%   \centering
%   \includegraphics{CA30PRG.pdf}
%   \caption{Evolution d'un automate cellulaire.}
%   \label{fig:CA30PRG}
% \end{figure}
Par convention, la règle décrite ci-dessus est appelée «règle 30», car
30 s'écrit 00011110 en binaire et 00011110 correspond à la deuxième
ligne du tableau.

Ces 256 automates cellulaires appelés également \emph{élémentaires}
ont été étudiés par Wolfram~\cite{wolfram} qui a proposé une
représentation alternative des règles qui peuvent être vues comme des
fonctions Booléennes à (au plus) 3 variables. Par exemple, la fonction
Booléenne qui correspond à la règle 30 est:
\begin{equation}\label{eq:ms30}
x_i^{t+1}=x_{i-1}^t\oplus (x_{i}^t\vee x_{i+1}^t)
\end{equation}
où $\oplus$ représente le «ou exclusif» et $\vee$ l'opération du «ou
logique».  Puisque nous nous intéressons essentiellement à la qualité
de l'aléatoire qu'on peut obtenir par l'évolution d'un automate
cellulaire, il est possible d'ex\-hiber des équivalences entre les
règles élémentaires. Nous en présentons trois: la \emph{conjonction},
la \emph{réflexion} et celle qui est la \emph{composition} des deux.
Remarquons cependant que le contenu de la cellule $i$ au temps $t$,
$x_i^t$ (obtenu de la configuration initiale $x^0$ par $t$ itérations
de la règle $r$) n'est pas le même que $(x'_i)^t$ (obtenu par $t$
itérations de la règle $r'\simeq r$) mais les comportements ont
globalement les mêmes propriétés.

Rappelons qu'une règle élémentaire est toujours de la forme $f(x)=y$
avec $x\in (\field_2)^3$ et $y\in\field_2$. Nous noterons dans la
suite $\tilde{m}$ le mot obtenu de $m=m_0m_1\ldots m_n$ en le lisant
de la droite vers la gauche: $\tilde{m}=m_n m_{n-1}\ldots m_0$ et
$\overline{m}$ le mot obtenu de $m$ en le complémentant bit à bit:
$\overline{m}=\overline{m}_0\ldots\overline{m}_n$.

\subsection{Conjonction}
\index{conjonction}
Soit la transformation:
$y_i=\overline{f(\overline{x_i})}\;i\in[\![0,7]\!]$ alors,
l'ens\-emble des entrées $x$ est transformé en:
{\small \[\begin{array}{|cccccccc|}
\hline\hline
0&1&2&3&4&5&6&7\\\hline 7&6&5&4&3&2&1&0\\\hline\hline
\end{array}\]}
il reste encore à complémenter les sorties pour obtenir la règle
équivalente par conjonction.
Par cette transformation, la règle 30 est transformée en 135.

Une autre fa\c{c}on de voir la règle $r'\simeq_{c}r$ écrite sur 8
bits est: $\tilde{r}=\overline{r'}$, donc $r'=\overline{\tilde{r}}$.

\subsection{Réflexion}
\index{réflexion}
Soit la transformation:
$y_i=f(\tilde{x_i})\;i\in[\![0,7]\!]$ alors, l'ens\-emble des
entrées $x$ est transformé en:
{\small \[\begin{array}{|cccccccc|}
\hline\hline
0&1&2&3&4&5&6&7\\\hline 0&4&2&6&1&5&3&7\\\hline\hline
\end{array}\]}
De cette manière, la règle 30 est transformée en 86.
\index{règle 86}

\subsection{Conjonction-réflexion}
\index{conjonction-réflexion}
On compose les deux transformations précédentes pour obtenir:
$y_i=\overline{f(\overline{\tilde{x_i}})}\;i\in[\![0,7]\!]$ alors,
l'ensemble des entrées $x$ est transformé en:
{\small \[\begin{array}{|cccccccc|}
\hline\hline
0&1&2&3&4&5&6&7\\\hline 7&3&5&1&6&2&4&0\\\hline\hline
\end{array}\]}
De cette manière, la règle 30 est transformée en 149.

\section{Le chiffre de Vernam}

Soit $P=p_1p_2\ldots p_m$ un clair de $m$ bits et $k_1k_2\ldots k_m$
un flux binaire; la clé $k$. Soit $c_i$ le $i$\ieme\ bit du chiffré
obtenu par application de l'opération de chiffrement $c_i=p_i\oplus
k_i$. Le déchiffrement est obtenu grâce à l'idempotence de $\oplus$ en
recalculant $p_i=c_i\oplus k_i$ avec la même clé $k$.

Ce chiffre, appelé chiffre de Vernam, est \emph{parfaitement
  sûr}~\cite{shannon,recif_00002046} pourvu que la clé soit une suite
aléatoire parfaite et qu'elle ne soit utilisée qu'une seule fois. D'un
point de vue pratique, pour utiliser ce chiffre, cela signifie qu'on
doit pouvoir construire assez facilement des suites
(pseudo)-aléatoires assez longues.

L'idée de Wolfram~\cite{wo-crypto} est d'utiliser un \ca pour
engendrer une telle suite pseudo-aléatoire en utilisant la suite des
valeurs prises au cours du temps par une cellule fixée avec un \ca
exécutant la règle 30 sur un anneau de $N$ cellules à partir d'une
configuration initiale qui joue le rôle de la clé du générateur de
suite pseudo-aléatoire. L'application itérée de la règle locale $f$
transforme celle-ci en une fonction Booléenne $F$ à $N$ variables qui,
au bout d'un certain temps, combine les valeurs de la configuration
initiale entre elles.

Dans les sections suivantes, nous verrons les faiblesses d'une telle
réalisation en rappelant tout d'abord une attaque menée contre le
générateur basé sur la règle 30. Nous montrerons ensuite en explorant
l'ensemble des règles élémentaires, qu'il n'existe pas de règle
élémentaire qui permette d'engendrer des suites pseudo-aléatoires de
bonne qualité cryptographique.

\section{L'attaque de Meier et Staffelbach}

Nous rappelons l'attaque de Meier et Staffelbach~\cite{ms91} à
clair/chiffré connus contre la règle 30.  Le but est de trouver un
chiffre équivalent qui ramène le problème de déduire la clé originale
à celui de trouver celle du chiffre équivalent. Le chiffre équivalent
possède un plus petit nombre de clés dont certaines ont une
plus grande probabilité d'apparition. Ces remarques permettent de
concevoir un algorithme d'attaque efficace.

La méthode est la suivante. On s'intéresse à la suite $\{x_i\}_t$ des
valeurs prises par la cellule $x_i$ étant donnée une
configuration initiale $S(t)=\{x_{i-n}^t,\ldots,x_i^t,\ldots,$
$x_{i+n}^t\}$, la clé. L'évolution de l'\ca est régi par la règle 30
qui peut être réécrite en utilisant la linéarité partielle:
\begin{equation}\label{eq:ms30inv}
x_{i-1}^t=x_{i}^{t+1}\oplus (x_i^t \vee x_{i+1}^t)
\end{equation}

Par l'équation~(\ref{eq:ms30}), les valeurs des cellules autour de
$x_i$ forment un triangle (cf. fig.~\ref{fig:triangle}).

\begin{figure}[th]
{\small \[\begin{array}{ccccccccccc}
x_{i-n}^t&\star&\ldots&\star&x_{i-1}^t&x_i^t&x_{i+1}^t&\star&\ldots&\star&x_{i+n}^t\\
\dummy&\star&\ldots&\star&x_{i-1}^{t+1}&x_i^{t+1}&x_{i+1}^{t+1}&\star&\ldots&\star&\dummy\\
\dummy&\dummy&\dummy&\star&\vdots&\vdots&\vdots&\star\\
\dummy&\dummy&\dummy&\dummy&\star&\star&\star\\
\dummy&\dummy&\dummy&\dummy&\dummy&x_i^{t+n}\\
\end{array}\]}
\caption{Triangle déterminé par la configuration initiale.}
\label{fig:triangle}
\end{figure}

Étant données les valeurs des cellules de deux colonnes adjacentes, la
{\em complétion arrière} permet de reconstruire par (\ref{eq:ms30inv})
la totalité du triangle à gauche de la suite $\{x_i\}_t$. Par la
complétion arrière, $N-1$ valeurs de $\{x_i\}_t$ et $N-2$ valeurs de
$\{x_{i+1}\}_t$ déterminent la configuration initiale. De manière
similaire, la clé peut être reconstruite à partir de $N-2$ valeurs de
$\{x_i\}_t$ et $N-1$ valeurs de $\{x_{i-1}\}_t$.

Si on se donne $N-1$ valeurs de la suite $\{x_i\}_t$, connaître la clé
équivaut à la connaissance d'une des suites adjacentes. Celles-ci
peuvent être considérées chacune comme une clé déterminant le reste de
la suite. Le problème devient: trouver une suite adjacente puis
déterminer la clé par complétion arrière.

Il y a certaines différences entre les suites adjacentes gauches et
droites que nous expliquons pour un \ca général de largeur $2n+1$ sans
condition de bord. On suppose connue $\{x_i\}_t^{t+n}$. Selon la
figure~\ref{fig:triangle}, le problème de trouver la suite adjacente
gauche est équivalent par (\ref{eq:ms30}) à celui de compléter la
totalité du triangle gauche. Réciproquement, la connaissance de
$L=\{x_{i-n}^t,\ldots,x_{i-1}^t\}$ et de $\{x_i\}_t$ équivaut à la
connaissance de tout le triangle gauche. Remarquons que les valeurs de
$L$ ne peuvent pas être choisies au hasard. Par exemple, si
$x_i^t=1\Rightarrow x_{i-1}^t=x_i^{t+1}+1$.

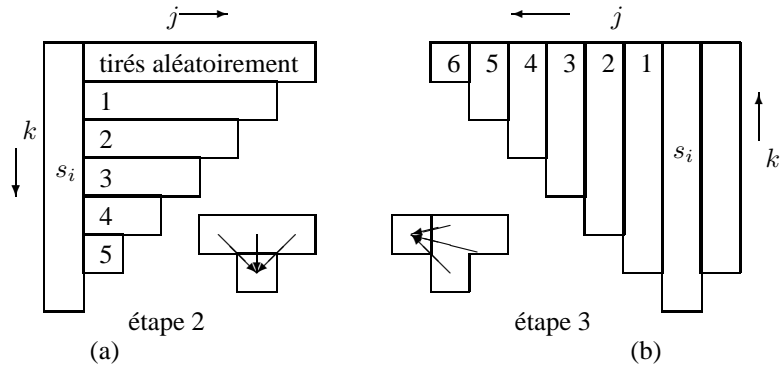
\begin{figure}[th]
\centerline{\setlength{\unitlength}{0.010000in}%
\begin{picture}(385,187)(5,645)
\thinlines
\put( 40,790){\framebox(120,20){}}
\put( 40,770){\framebox(100,20){}}
\put( 40,750){\framebox(80,20){}}
\put( 40,730){\framebox(60,20){}}
\put( 40,710){\framebox(40,20){}}
\put( 40,690){\framebox(20,20){}}
\put( 90,825){\vector( 1, 0){ 25}}
\put(  5,755){\vector( 0,-1){ 25}}
\put(340,670){\framebox(20,140){}}
\put(360,690){\framebox(20,120){}}
\put(320,690){\framebox(20,120){}}
\put(280,730){\framebox(20,80){}}
\put(260,750){\framebox(20,60){}}
\put(220,790){\framebox(20,20){}}
\put(240,770){\framebox(20,40){}}
\put(300,710){\framebox(20,100){}}
\put(292,825){\vector(-1, 0){ 30}}
\put(  5,755){\vector( 0,-1){ 25}}
\put(390,759){\vector( 0, 1){ 25}}
\put( 20,670){\framebox(20,140){}}
\put(100,700){\framebox(60,20){}}
\put(120,680){\framebox(20,20){}}
\put(110,710){\vector( 1,-1){ 20}}
\put(130,710){\vector( 0,-1){ 20}}
\put(150,710){\vector(-1,-1){ 20}}
\put(260,720){\line(-1, 0){ 40}}
\put(220,720){\line( 0,-1){ 40}}
\put(220,680){\line( 1, 0){ 20}}
\put(240,680){\line( 0, 1){ 20}}
\put(240,700){\line( 1, 0){ 20}}
\put(260,700){\line( 0, 1){ 20}}
\put(260,720){\line( 0, 1){  0}}
\put(200,700){\framebox(20,20){}}
\put(230,690){\vector(-1, 1){ 20}}
\put(230,715){\vector(-4,-1){ 20}}
\put(244,701){\vector(-4, 1){ 34.118}}
\put( 45,795){\makebox(0,0)[lb]{\raisebox{0pt}[0pt][0pt]{ tir\'es al\'eatoirement}}}
\put( 45,695){\makebox(0,0)[lb]{\raisebox{0pt}[0pt][0pt]{ 5}}}
\put( 45,715){\makebox(0,0)[lb]{\raisebox{0pt}[0pt][0pt]{ 4}}}
\put( 45,735){\makebox(0,0)[lb]{\raisebox{0pt}[0pt][0pt]{ 3}}}
\put( 45,755){\makebox(0,0)[lb]{\raisebox{0pt}[0pt][0pt]{ 2}}}
\put( 80,820){\makebox(0,0)[lb]{\raisebox{0pt}[0pt][0pt]{ $j$}}}
\put(  5,760){\makebox(0,0)[lb]{\raisebox{0pt}[0pt][0pt]{ $k$}}}
\put(310,820){\makebox(0,0)[lb]{\raisebox{0pt}[0pt][0pt]{ $j$}}}
\put(390,745){\makebox(0,0)[lb]{\raisebox{0pt}[0pt][0pt]{ $k$}}}
\put(342,750){\makebox(0,0)[lb]{\raisebox{0pt}[0pt][0pt]{ $s_i$}}}
\put( 45,775){\makebox(0,0)[lb]{\raisebox{0pt}[0pt][0pt]{ 1}}}
\put(325,795){\makebox(0,0)[lb]{\raisebox{0pt}[0pt][0pt]{ 1}}}
\put(305,795){\makebox(0,0)[lb]{\raisebox{0pt}[0pt][0pt]{ 2}}}
\put(285,795){\makebox(0,0)[lb]{\raisebox{0pt}[0pt][0pt]{ 3}}}
\put(265,795){\makebox(0,0)[lb]{\raisebox{0pt}[0pt][0pt]{ 4}}}
\put(245,795){\makebox(0,0)[lb]{\raisebox{0pt}[0pt][0pt]{ 5}}}
\put(225,795){\makebox(0,0)[lb]{\raisebox{0pt}[0pt][0pt]{ 6}}}
\put( 60,660){\makebox(0,0)[lb]{\raisebox{0pt}[0pt][0pt]{ \'etape 2}}}
\put(260,660){\makebox(0,0)[lb]{\raisebox{0pt}[0pt][0pt]{ \'etape 3}}}
\put( 40,645){\makebox(0,0)[lb]{\raisebox{0pt}[0pt][0pt]{ (a)}}}
\put(320,645){\makebox(0,0)[lb]{\raisebox{0pt}[0pt][0pt]{ (b)}}}
\put( 22,740){\makebox(0,0)[lb]{\raisebox{0pt}[0pt][0pt]{ $s_i$}}}
\end{picture}}
\caption{Explication du fonctionnement de l'algorithme. En (a),
on illustre la complétion avant et en (b) la complétion arrière.
On peut aussi voir le fonctionnement de l'algorithme comme si on
appliquait le masque de droite dans le sens $j,k$ pour (a) et $k,j$ pour (b).}
\label{fig:explique}
\end{figure}

D'un autre coté, tout choix de $R=\{x_{i+1}^t,\ldots,x_{i+n}^t\}$
mène à une complétion consistante vis à vis de $\{x_i\}_t$.
En effet, par (\ref{eq:ms30inv}), pour tout
élément de la suite adjacente droite $x_{i+1}^t$ il existe un
élément de la suite adjacente gauche $x_{i-1}^t$ consistant avec
la valeur suivante de la suite de référence $x_i^{t+1}$. De plus,
selon (\ref{eq:ms30}), n'importe quel choix de
$\{x_{i+2}^t,\ldots,x_{i+n}^t\}$ mène à une extension
consistante en $x_{i+2}^{t+1},\ldots,x_{i+n-1}^{t+1}$. En itérant ce
procédé, on obtient le triangle de droite de la
figure~\ref{fig:triangle}. Ce procédé, appelé {\em complétion
avant}, construit une suite adjacente droite consistante avec la suite
de référence pour tout choix de $R$.

Supposons à présent que l'\ca est un anneau de $N$ cel\-lules (ce qui
signifie en particluier que $L=R$).  Pour retrouver la clé, on peut
calculer soit $R=\{x_{i}^t,\ldots,x_{i+N-1}^t\}$, soit
$L=\{x_{i-N+1}^t,\ldots,x_i^t\}$. En s'appuyant sur les remarques
précédentes, on obtient l'algorithme MS, illustré par la
figure~\ref{fig:explique}.

\begin{center}
\begin{minipage}[htb]{12cm}
\paragraph{Algorithme MS}
{\small \begin{enumerate}
\item engendrer une clé aléatoire;
\item complétion avant:
\begin{tabbing}
{\bf Pour }\= $k\in$ \= $\{1,\ldots,N-2\}$\\
\>{\bf Pour }$j\in\{1,\ldots,N-k-1\}$\\
\>\>$x_{i+j}(t+k)\leftarrow x_{i+j-1}(t+k-1)\oplus (x_{i+j}(t+k-1)\vee x_{i+j+1}(t+k-1))$;
\end{tabbing}
\item complétion arrière:
\begin{tabbing}
{\bf Pour }\= $j\in$ \= $\{1,\ldots,N-1\}$\\
\>{\bf Pour }$k\in\{N-1-j,\ldots,0\}$\\
\>\>$x_{i-j}(t+k)\leftarrow x_{i-j+1}(t+k+1)\oplus (x_{i-j+1}(t+k)\vee x_{i-j+2}(t+k))$;
\end{tabbing}
\item exécuter la règle 30 sur la config. init. obtenue ci-dessus pour générer
la suite. Fin si co\"{\i}ncidence, sinon aller en 1.
\end{enumerate}}
\end{minipage}
\end{center}

Le fonctionnement de l'algorithme est illustré sur un exemple avec
$N=5$ par la fig.~\ref{fig:exemple attaque ms}. Dans ce cas, la clé
est $(x_{i},\ldots,x_{i+4})=(0,1,0,1,1)$ et la suite pseudo-aléatoire
$(0,0,1,0,0)$. A l'étape~1 de l'algorithme, les valeurs
$x_{i+1},\ldots,x_{i+4}$ sont choisies au hasard. Si $x_{i+1}$ est à
1, alors par (\ref{eq:ms30}), la suite adjacente droite produite est
indépendante du choix de $x_{i+2},x_{i+3}$ et $x_{i+4}$. C'est pour
cela qu'ils sont marqués par «$\star$» dans la fig.~\ref{fig:exemple
  attaque ms}.  Ainsi, il n'y a qu'une seule suite adjacente droite
avec $x_{i+1}=1$.  Donc, avec probabilité $1/2$, la clé correcte est
trouvée dès le premier essai.

\begin{figure}[th]
{\small\[\begin{array}{ccccccccc}
x_{i-4}&x_{i-3}&x_{i-2}&x_{i-1}&x_{i}&x_{i+1}&x_{i+2}&x_{i+3}&x_{i+4}\\
1&0&1&1&\fbox{{\bf 0}}&\fbox{1}&\fbox{0}&\fbox{1}&\fbox{1}\\
\dummy&0&1&0&{\bf 0}&1&0&1\\
\dummy&\dummy&1&1&{\bf 1}&1&0\\
\dummy&\dummy&\dummy&0&{\bf 0}&0\\
\dummy&\dummy&\dummy&\dummy&{\bf 0}\\
\end{array}\]}
\[\mbox{\it Génération de la suite temporelle}\]
{\small\[\begin{array}{ccccccccc}
x_{i-4}&x_{i-3}&x_{i-2}&x_{i-1}&x_{i}&x_{i+1}&x_{i+2}&x_{i+3}&x_{i+4}\\
\fbox{1}&\fbox{0}&\fbox{1}&\fbox{1}&\fbox{{\bf 0}}&1&\star&\star&\star\\
\dummy&0&1&0&{\bf 0}&1&\star&\star\\
\dummy&\dummy&1&1&{\bf 1}&1&\star\\
\dummy&\dummy&\dummy&0&{\bf 0}&0\\
\dummy&\dummy&\dummy&\dummy&{\bf 0}\\
\end{array}\]}
\vspace*{-.2cm}
\[\mbox{\it Détermination de la racine par complétion arrière}\]
\caption{Un exemple simple de cryptanalyse.}
\label{fig:exemple attaque ms}
\vspace*{-.2cm}
\end{figure}
Puisque la règle 30 est susceptible d'une attaque, qu'en est-il des
autres règles? Existe-t'il, parmi les 256 règles élémentaires,
de meilleures règles que celle proposée par Wolfram? C'est à cette
question que nous allons répondre dans la section suivante.

\section{Étude de l'auto-corrélation}

Une mesure du taux de corrélation entre les entrées et les sorties des
itérées des fonctions de transition des automates cellulaires
élémentaires peut être obtenue à l'aide de la transformée de Walsh.

Notons $F(\underline{x})$ la valeur de la fonction $F$ évaluée au
point défini par le vecteur
$\underline{x}=(x_0,x_1,\ldots,x_{N-1})$ de $(\field_2)^N$ ou, de
manière équivalente, $F(x)$ la valeur de $F$ évaluée au point
$x=\sum_{i=0}^{N-1}x_i.2^i$. Soit également
$\underline{\omega}\in(\field_2)^N$ et
$\omega=\sum_{i=0}^{N-1}\omega_i.2^i$, l'entier correspondant.
On définit la {\em transformée de Walsh} de $F$ par $\hat{F}
(\omega)=\sum_{x=0}^{2^N-1}F(x)(-1)^{\langle\underline{x},\underline{\omega}\rangle}$,
où $\langle\underline{x},\underline{\omega}\rangle$ est le produit
scalaire.

La transformée de Walsh possède quelques propriétés statistiques
intéressantes: la valeur de la transformée au point 0 est égal à la
valeur moyenne de la fonction: $\hat{F}(0)=E[F(x)]$. Cette propriété
permet de tester l'équidistribution de 0 et de 1 de $F$. En effet, si
$F$ est équidistribuée, $F(0)=2^{N-1}$.

On peut aussi déduire des dépendances statistiques entre des
sous-ensembles des entrés et la sortie de $F$. Il est clair que si on
connait le vecteur d'entrée complet de $F$, la valeur de la sortie est
sans ambigu\"{\i}té aucune.  Mais il peut y avoir certaines variables
d'entrée ou de petits sous-ensembles de variables d'entrée qui
réduisent l'incertitude sur la sortie de $F$. En d'autres termes, il y
a une corrélation entre tout sous-ensemble $X_{1},\ldots,X_{m}$ de
variables d'entrée et la variable de sortie $Z$ de $F$. C'est ce
qu'énonce le lemme~\ref{lem:rueppel}:
\begin{lemma}(Rueppel~\cite{rueppel})\label{lem:rueppel}
La variable aléatoire discrète $Z$ est indépendante des $m$
variables aléatoires indépendantes et uniformément distribuées
$X_1,\ldots,X_m$ \ssi $Z$ est indépendante de la somme sur
$\field_2$ de $\sum_{i=1}^{m}c_iX_i$ pour tout choix de
$c_1,c_2,\ldots,c_m$ non tous nuls de $\field_2$.
\end{lemma}

Ce lemme implique que l'information mutuelle entre la variable de
sortie et les $m$ variables d'entrée est nulle, \ssi l'information
mutuelle entre $Z$ et toute combinaison linéaire non nulle de $m$
variables d'entrée vaut zéro. On en déduit:
$P[F=1/\langle X,\omega\rangle =1]=\frac{1}{2}-\frac{F(\omega)}{2^N}$
\begin{lemma}
  Soit $X=(X_1,\ldots,X_m)$ un vecteur aléatoire où les V.A. binaires
  $X_1,\ldots,X_m$ sont indépendantes et telles que
  $P[X_i=0]=P[X_i=1]$ pour tout $i$. Soit
  $F:(\field_2)^N\rightarrow\field_2$ et soit $\omega\neq 0$. Alors,
  la V.A. $Z=F(X)$ est indépendante de $\langle\omega,X\rangle$ \ssi
  $\hat{F}(\omega)=0$.
\end{lemma}
Nous pouvons maintenant énoncer le critère principal de résistance aux
corrélations:
\begin{theorem}[Xiao et Massey~\cite{XiaoM88}]
  La fonction $F:(\field_2)^N\rightarrow\field_2$ est résitante aux
  corrélations à l'ordre $k$ \ssi $\hat{F}(\omega)=0,$
  $\forall\omega=(\omega_1,\ldots,\omega_t)\neq 0$ tel que le nombre de
  $\omega_i$ non nuls est au plus $k$.
\end{theorem}

En fait, l'idée d'utiliser la transformée de Walsh pour tester la
qualité d'une suite pseudo-aléatoire vient
de~\cite{Yuen77c}. Dans cet article, Yuen a observé
que le spectre d'une suite parfaitement aléatoire est asymptotiquement
plat. Cette observation a été ensuite utilisée pour construire des
tests pour mesurer la qualité des suites pseudo-aléatoire en
améliorant ceux proposés traditionnellement par Knuth~\cite{knut}.

Nous avons utilisé ces différentes propriétes pour trouver, parmi les
règles élémentaires, celles qui sont les plus susceptibles d'engendrer de
l'aléatoire d'une qualité convenable~\cite{hiroshima2006}. Pour appliquer
ces différentes propriétés, nous allons calculer la transformée de
Walsh en utilisant un algorithme proposé dans \cite{elliott}.

Nous procédons en plusieurs étapes. La première est de
rechercher parmi toutes les règles élémentaires celles qui sont
équidistribuées en calculant la transformée de Walsh de chaque
règle et en sélectionnant les règles $i$ telles que
$\hat{F_i}(0)=4$. Ce premier tri nous permet de retenir 70 règles.

La seconde est de rechercher, parmi ces 70 règles,
les meilleures règles. Pour ce faire, nous nous
intéressons aux itérées des fonctions sélectionnées et nous
choisirons la (ou les) fonction(s) $F_i$ vérifiant:
\[\min_{F_i}\max_{\omega=2^k} |\widehat{F^{(o)}_i}(\omega)|\]
o\`u $o$ représente le nombre d'itérations de $F_i$ et
en s'intéressant aux $\omega$ de la forme $2^k, k\in[\![ 0,2.o+1]\!]$.

La recherche exhaustive des meilleures fonctions a été menée en
calculant la transformatée de Walsh sur toutes les règles jusqu'à la
cinquième itérée ($o=5$) avec un algorithme dont la complexité globale
est asymptotiquement: $256.o^2.(2.o+1).2^{2.o+1}$. Les règles
intéressantes sont reportées dans le tableau~\ref{table:result} dans
lequel toutes les fonctions de valeur 0 sont exactement des règles
correspondant à des fonctions linéaires. Les seules bonnes fonctions
non linéaires peuvent être obtenues de la règle 30 par le biais des
équivalences que nous avons rappelées dans le tableau. D'où:
\begin{theorem}
  Il n'existe pas de règle élémentaire non linéaire d'automate
  cellulaire qui soit résistante aux corrélations.
\end{theorem}

On retrouve ainsi le fait que la règle 30 (ainsi que les règles qui
lui sont équivalentes par les transformations) est une «bonne» règle
pour la génération de suites pseudo-aléatoires. Cependant, nous avons
vu que cette règle est peu robuste à la cryptanalyse inventée par
Meier et Staffelbach. Dans la prochaine section, nous donnons quelques
pistes pour construire de meilleurs automates cellulaires pour
engendrer des suites pseudo-aléatoires.

\begin{table}
\scriptsize{
\[\begin{array}{|r||r|r||r|r||r|r||r|r||r|r||r|r|r||}
\hline\hline
\mbox{ordre}&1&\dummy&2&\dummy&3&\dummy&4&\dummy&5&\dummy&\mbox{conj}&\mbox{refl}&\mbox{c.r.}\\
\mbox{règle}&\mbox{cfg}&\mbox{val}&\mbox{cfg}&\mbox{val}&\mbox{cfg}&\mbox{val}&\mbox{cfg}&\mbox{val}&\mbox{cfg}&\mbox{val}&\dummy&\dummy&\dummy\\\hline
30&4&2&16&4&64&16&256&40&1024&80&135&86&149\\
60&0&0&0&0&0&0&0&0&0&0&195&102&153\\
86&1&2&1&4&1&16&1&40&1&80&149&30&135\\
90&0&0&0&0&0&0&0&0&0&0&165&90&165\\
102&0&0&0&0&0&0&0&0&0&0&153&60&195\\
105&0&0&0&0&0&0&0&0&0&0&105&105&105\\
135&4&2&16&4&64&16&256&40&1024&80&30&149&86\\
149&1&2&1&4&1&16&1&40&1&80&86&135&30\\
150&0&0&0&0&0&0&0&0&0&0&150&150&150\\
153&0&0&0&0&0&0&0&0&0&0&102&195&60\\
165&0&0&0&0&0&0&0&0&0&0&90&165&90\\
195&0&0&0&0&0&0&0&0&0&0&60&153&102\\
\hline\hline
\end{array}\]}
\caption{Résultat de la transformée de Walsh et ``bonnes''
  fonctions. Dans ce tableau, l'ordre correspond au nombre
  d'itérations de la règle, val la plus petite valeur de la
  transformée de Walsh et cfg la configuration correspondante. A
  droite, nous avons rappelé les règles équivalentes.}
\label{table:result}
\end{table}

\section{De nouvelles pistes}
Dans~\cite{sipper96coevolving}, Tomassini et Sipper ont suggéré
l'utilisation d'\cas non uniformes pour engendrer de meilleures suites
pseudo-aléatoires. Dans ce modèle, chaque cellule peut utiliser
plusieurs règles (l'\ca devient \emph{non-uniforme}) et les meilleures
règles sont sélectionnées par une approche évolutionnaire au moyen
d'un algorithme génétique. De cette manière, Tomassini et Sipper ont
sélectionné quatre règles de rayon un; il s'agit des règles  $90$,
$105$, $150$ and $165$ qui sont toutes linéaires, ce qui est un
inconvénient, comme il est rappelé dans~\cite{zemor}. Ils ont utilisé
une batterie de tests statistiques pour mesurer la qualité des suites
pseudo-aléatoires engendrées avec de bons résultats.

Cette étude a été généralisée à des règles de rayon supérieur à un
dans~\cite{1016320}. Seredynski \emph{et al} ont à leur tour proposé
différentes règles de rayon 1 et 2 d'automates cellulaires; ce sont
les règles de rayon un $30$, $86$ et $101$ et les règles de rayon deux
$869020563$, $1047380370$, $1436194405$, $1436965290$, $1705400746$,
$1815843780$, $2084275140$ et $2592765285$. Leur nouvel ensemble de
règles a été testé selon les directives du FIPS 140-2~\cite{FIPS}
et des tests de Marsaglia, implémentés dans le paquetage
\texttt{diehard}.

Cependant, dans aucun des deux cas, les auteurs n'ont mené une étude
de la corrélation des suites engendrées. Il faudrait faire une étude
analogue à la notre pour valider l'approche évolutionnaire de
sélection des règles d'\cas pour la génération de suites
pseudo-aléatoires.

\bibliographystyle{plain}
\bibliography{toute,mb}
\end{document}